# SAR Focused Microwave Reflection Tomography for Biomedical Imaging

Kevin K.M. Chan, Daniel Oloumi, Pierre Boulanger and Karumudi Rambabu

*Abstract*—This paper proposes the combination of SAR and microwave reflection tomography for biomedical imaging applications. The proposed method can achieve accuracies of less than 5mm and does not suffer from instability during reconstruction. Fourier based reconstruction is adopted which makes it fast and computationally efficient.

## I. INTRODUCTION

Biomedical imaging involves the representation of biological tissues in the spatial domain for assessing their condition without surgery which minimizes risk to a patient. Mainstream bio-imaging modalities include x-ray, computed tomography (CT) which is a combination of multiple x-ray scans to produce a slice/volume image, emission computed tomography (ECT) which involves the use of radioactive material as tracers for illuminating high concentration areas of particular compounds within living tissues, ultrasound and magnetic resonance imaging (MRI) [1]. The state-of-the-art method that produces the best image is the MRI having resolution in the order of millimeters, though it is more suited for imaging soft tissues due to their higher concentration of the hydrogen molecules. The disadvantages of MRI are its cost, system complexity and time needed for data acquisition where the patient must stay still for periods of up to an hour in the MRI chamber. Ultrasound on the other hand is a low-cost solution, has low complexity and can be portable. The down side of ultrasound is that it requires a good physical contact with the imaged site and a high sonic wave impedance contrast for a good image. Remote sensing x-ray, CT and ECT modalities expose the patients to ionizing radiation that increases the risk of cell mutation.

Microwave imaging is an alternative modality that has great potential for biomedical applications and beyond. Using non-ionizing microwaves with emission power levels typically less than cellular device radiations, while being non-invasive and having system complexity similar to ultrasound technology, microwave imaging has garnered lots of recent attention among research groups to develop solutions.

The common approach among the microwave community in developing a biomedical imaging solution is through the inverse scattering method [2-3]. For inverse scattering, measured fields depicting the object function are back-propagated for image reconstruction. Oftentimes, the systems are narrow band and typical measurement schemes employ the use of vector network analyzers. Multi-frequency measurements tend to give better reconstructions as the Fourier space is filled to a fuller extent. Back-scattering (reflection) rather than forward-scattering (transmission) measurement is more viable for biomedical applications because the organic tissues are generally electro-magnetically lossy due to high water content. However, for back-scattered measurements, the reconstruction algorithm using back-propagation is unstable and often leads to a non-conclusive solution [4].

Inverse scattering is a non-linear problem where linearization using method such as Born or Rytov approximation is commonly applied. For Born approximation, weak scattering is assumed where the impedance contrast between the incident and scattered fields should not exceed 20% [5]. Many living organs have high dielectric constants thus the weak scattering condition for Born is compromised that result in a lesser accuracy. Many groups use matching fluid to control the dielectric contrast ratio such that the weak scattering condition is valid. Iterative methods that use optimization routines such as Gauss–Newton and conjugate gradient are alternatives that can solve high dielectric contrast in biomedical applications, at the expense of computational load and the risk of instability.

Recent examples of microwave tomography system in medical imaging include the use of parallel computing via domain decomposition to speed up the reconstruction process [6] and the use of typical dielectric constants for biological tissues as a priori settings applied to a modified Gauss-Newton inversion to enhance the reconstruction [7]. Improvements are generally focused on the inversion process for higher speed or better accuracy.

Wideband microwave measurements, such as pulse-based ultra-wideband (UWB) radars are not common place in biomedical imaging solutions. Space-time beamforming [8] and confocal [9] UWB radar medical imaging systems are approaches to obtain images via time-domain pulses, however they are not tomographic. It is advantageous to use UWB radar as the system is analogous to ultra-sound technology while being contactless. Ultra-sound is a back-scattering method which suits the measurement scheme for microwaves when applied to biomedical. Typical UWB radar operates in the frequency range of 3 to 10 GHz. Majority of the pulse power is centered around the 6 to 7 GHz frequency band. The free-space wavelength where the dominant energy is focused is ~4cm. For high dielectric constant materials, the wavelength decreases further by a factor or √$\varepsilon_r$. Electro-magnetic scattering suffers from diffraction for object sizes that are in the order of its wavelength. Thus, diffraction effects for UWB

K.K.M. Chan, D. Oloumi and K. Rambabu are with the Electrical and Computer Engineering Department, University of Alberta, Edmonton, AB T6G1H9, Canada (e-mail: kkchan@ualberta.ca), and P. Boulanger is with the Computing Science Department, University of Alberta, Edmonton, AB T6G2E8, Canada.

band of frequency will be present for objects smaller than 4cm. Numerous bones have cross-sectional dimensions in that order and will diffract the incident microwaves.

The image reconstruction strategy for ultra-sound uses reflection tomography where an array of transducers detects echoes back-scattered from objects that represents a one-dimension projection. Computed tomography algorithm can be applied in this case as the projection is a form of straight line representation of the imaging plane. Using Fourier based CT algorithm is inherently stable and computationally efficient.

The measurement using a linear array configuration in reflection tomography is compatible with synthetic aperture radar (SAR) processing, When SAR processing is applied first from the array data, a focused image will result that addresses the edge diffractions from small objects. After SAR focusing, the resultant data can then be further processed using reflection tomography algorithm to yield the reconstructed object image.

This manuscript describes the results of microwave reflection tomography for a biological phantom model with SAR applied to mitigate the diffraction effects.

## II. MICROWAVE REFLECTION TOMOGRAPHY EXPERIMENTAL SETUP

Reflection tomography is the method of using the summation of B-mode scan data to form a 1-D projection of the imaged object [10]. Fig. 1 shows the scenario for a microwave reflection tomography measurement. A wideband pulse is incident on the imaging domain as a plane wave. This incident pulse can be radiated from a source placed at least the far-field distance away, for it to be considered as a plane wave. Alternatively, the same pulse can be transmitted at every receiver location to simulate the plane wave condition.

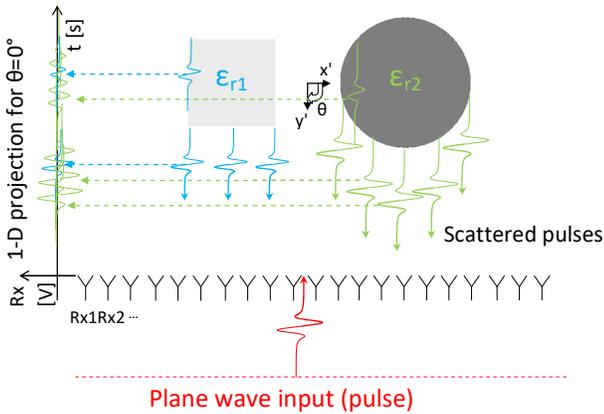

Fig. 1. Set-up diagram for microwave reflection tomography.

A linear array of receivers, which can be either a separate receiver at each location or a single receiver that scans the scene at each location (i.e. moves from one location to another with synchronization to the transmitted pulses, e.g. time alignment at each receiver location), is placed at a far-field distance away from the extent of the objects when rotated at all angles. The maximum distance between each receiver location is governed by the Nyquist sampling criterion for the given signal bandwidth, which translates to a pulse-spatial width for pulsed broadband systems. The plane wave input pulse applied in our work has a bandwidth from 1 to 10 GHz which results in a pulse-spatial width of 3cm. For our measurement set-up, an inter-spacing of 2cm is used.

From Fig. 1, the pulse reflections (illustrative) from the objects in the imaging domain due to dielectric contrast with respect to the bounding medium will be recorded by the array of receivers as a collection of A-mode scans at their given locations (forming a B-mode image). This scanning strategy is analogous to the ground penetrating radar system. When time aligned, as depicted in the plot on the left of Fig. 1, a 1-D projection of the imaging scene is captured when all the receiver data are summed together (e.g. the locations of medium boundaries are stated by their respective time indexes and the dielectric contrasts are captured through the pulse amplitudes and shapes). Multiple 1-D projections are measured as the objects are rotated by a pre-determined step angle or by traversing the source and linear array of receivers by an arc length due to the step angle. Then multiple angular 1-D projection data are used for the reconstruction process, like in computed tomography.

The axis of rotation, given by the x' and y' axes of Fig. 1, must be noted as its time index in the summed data will be at the middle of the time frame for all step angles. This can be done through a calibration process where a large metal plate is placed at the plane of rotation and the reflected pulses recorded such that the time index at that location (which will be the same for all Rx) is set as the middle of the time frame.

As reflection data is collected, and it is assumed that the dielectric contrast between the bounding medium (air) and the objects (biological tissues) is high such that little energy will penetrate through, a 360° scan with finite steps is required. The maximum angle of rotation per step is also determined by the Nyquist criterion, where the length of the circumferential arc due to the step angle (with the radius being the length from the axis of rotation to the extent of the objects) should be less than the system's pulse-spatial width.

Fig. 2 shows the test scenario for our proof of concept. Two objects are placed in the imaging domain with their physical dimensions correspond to the x' and y' axes scales. The objects are simplified shapes to depict a mid-thorax cross-section of the liver (left square with 10cm sides) and heart (right circle with 15cm diameter). The maximum extent of the radius containing all objects from the origin (0,0) is 17.5cm. For our system's pulse-spatial width of 3cm, the maximum step angle that produces a circumferential arc less than 3cm with a 17.5cm radius is 9.8°. A step angle of 5° is applied such that we will have 72 projections.

The phantom model of Fig. 2 is developed in CST Microwave Studio (full-wave transient electromagnetic solver) with the shapes extended in both directions in the z' axis by 30cm and are terminated with PML boundaries (e.g. only 2-D variation). Simulations are used for the proof of concept because all the dimensions are accurate and there is no time jitter in the measurements (i.e. all parameters are controlled and repeatable). The material properties for the simulated liver (square) are set at $\varepsilon_r$=37.8 and $\mu_r$=1, and the heart (circle) with $\varepsilon_r$=47.7 and $\mu_r$=1. The material data are taken from IT'IS Foundation's Tissue Properties for the frequency of 6.5GHz. The receivers are E-field probes placed 40cm away from the radius of the objects extent. 25 probes spaced 2cm apart measured the reflections for a 48cm span. The default transient source pulse with frequency from 1 to 10GHz is used as a plane wave excitation with incident direction towards the objects. The E-field vectors for the source pulse and receiver probes are linearly polarized and aligned to the z' axis.

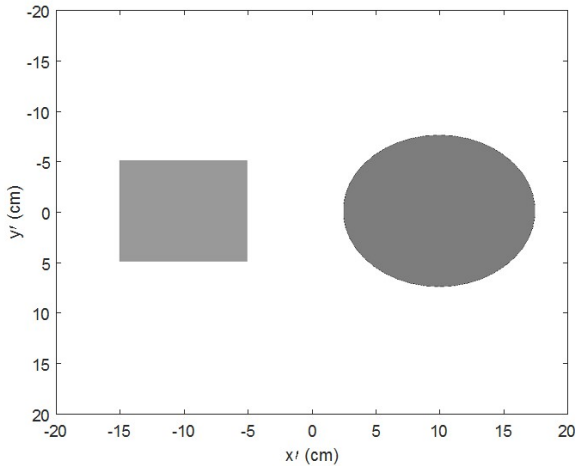
Fig. 2. Phantom diagram and dimensions.

## III. Data Processing

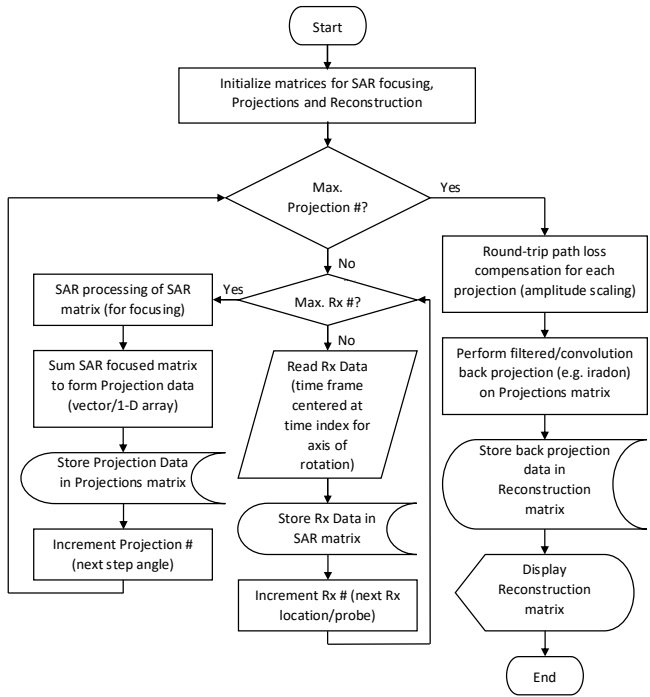
Fig. 3. Flowchart for the SAR focused reflection tomography algorithm.

Fig. 3 outlines the data processing methodology for our proposed SAR focused microwave reflection tomography system. When the receiver data are being scanned, the received data will contain edge diffractions for object with size in the order of the pulse-spatial width (range resolution of the radar system). The 1-D linear array arrangement (in x' axis) and data format (spatio-temporal relation) of the receivers are compatible with SAR processing. Therefore, to cause the scattering to "focus" in space (by synthesizing a larger antenna aperture), we first perform a SAR processing on the received data set for each angle of projection. The SAR algorithm performed is similar to the method presented in [11].

The SAR focused data (stored in a matrix) will then be summed to form a 1-D array which constitutes the projection data at the given angle.

The angle is incremented by the step size and a new set of receiver data collected. The SAR focusing followed by summing process are repeated for all step angles.

From the projections data, the object scene has been captured with edge diffractions minimized through SAR focusing. Round-trip path loss for each projection is first compensated for amplitude scaling. To reconstruct the scene, we perform the filtered/convolution back projection, similar to the computed tomography. The algorithm is stable and is not affected by convergence issue like in inverse scattering.

## IV. Results

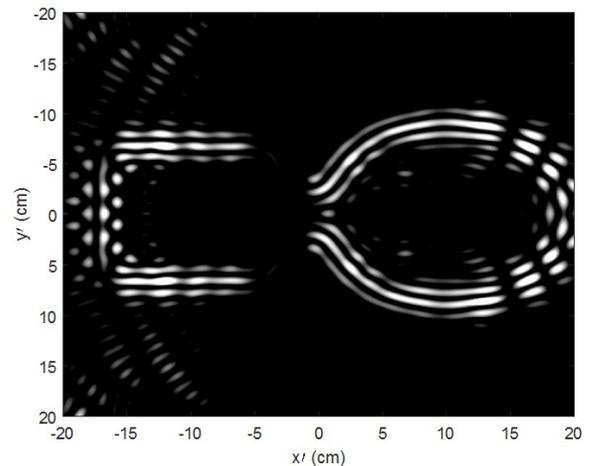
Fig. 4. Phantom reconstruction using reflection tomography.

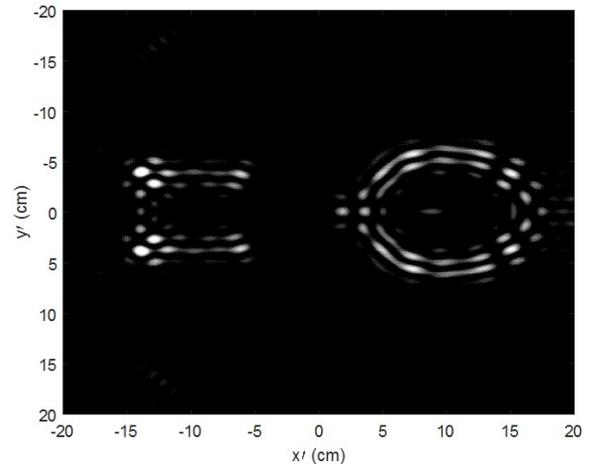
Fig. 5. Phantom reconstruction with addition of SAR.

Figs. 4 and 5 show the reconstructions for regular microwave reflection tomography and with SAR focusing applied respectively. From the observed thickness of the band, ~3cm for all edges (for both Fig. 4 and 5), the range resolution of our radar system having 9GHz bandwidth can be deduced. Fig. 4 contains more spreading of energy at the edges as compared to Fig. 5, which shows the benefit of applying SAR focusing. The estimated accuracy for the SAR focused image, when compared with the original phantom model of Fig. 2, is within 5mm.

## V. Application to Biomedical Imaging

To find out the performance of the proposed SAR focused microwave reflection tomography in a more realistic scenario as compared to the phantom model, we repeat the measurements and data processing using a biological voxel model. The adopted biological voxel model is Gustav, a 38-year-old adult male, from the CST voxel family. The model is developed from MRI/CT data and has a resolution of about 2mm. The tissues are segmented, and we have selected the thorax region for imaging. The entire thorax of 43cm height is included in the simulation. Only selected tissues are made active in the model. They include the bones, lungs, liver, heart and kidneys. Material properties are as defined in the voxel model in CST. Fig. 6 shows the axial cut of the thorax at the height of the linear receiver array. It is used as the reference for the image reconstruction.

The receiver array (E-field probes) is placed at 40cm from the extent of the voxel model and has an 80cm span. Spacings between the probes are maintained at 2cm as the same signal bandwidth is applied. A step angle of 5° is also adopted which is sufficient to meet the Nyquist criterion.

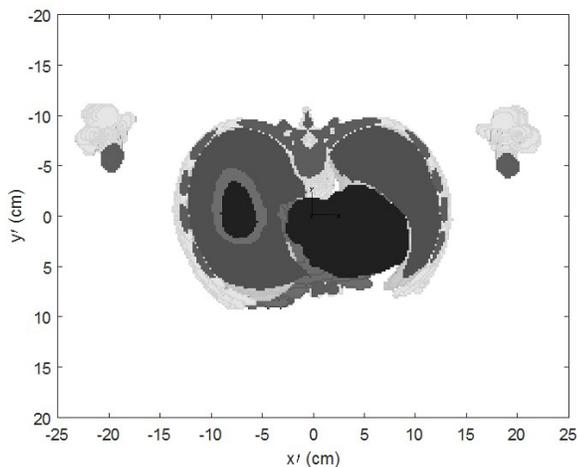

Fig. 6. Biological voxel model with dimensions.

Figs. 7 and 8 show the reconstructions of the voxel model data processed using microwave reflection tomography and with SAR focusing applied respectively. As the model is not homogenous in the z' axis, the reconstructions contain energy scattered from the other heights (z' axis locations) as the SAR focusing is applied only for the x' axis.

From Fig. 7, it can be observed that there is smearing of the image around the location of the bones. The dimensions of the bones are around or less than the spatial-pulse width of 3cm, and therefore will diffract the microwaves. The soft tissues are lossy, so the microwave energy is being absorbed in them. The main scatterers are therefore the bones.

The reconstructed image of Fig. 8 shows the outline of the bone structure quite well. Though not very well focused, the data does contain energy from the other heights and this can be observed predominantly for the limb bones, where the outlines of the elbow joints are captured in the image. The spinal cord outline is also reconstructed to within cm accuracy.

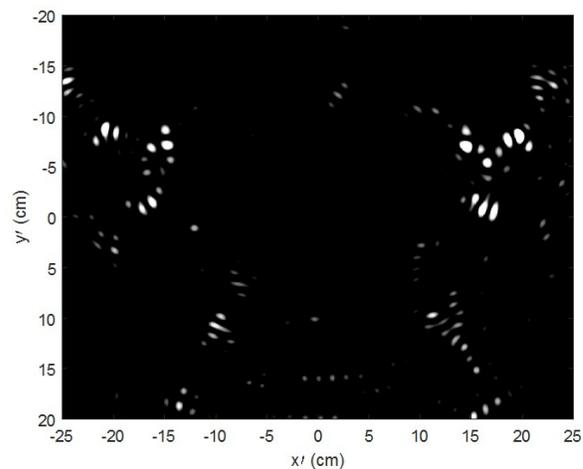

Fig. 7. Biological voxel model reconstruction using reflection tomography.

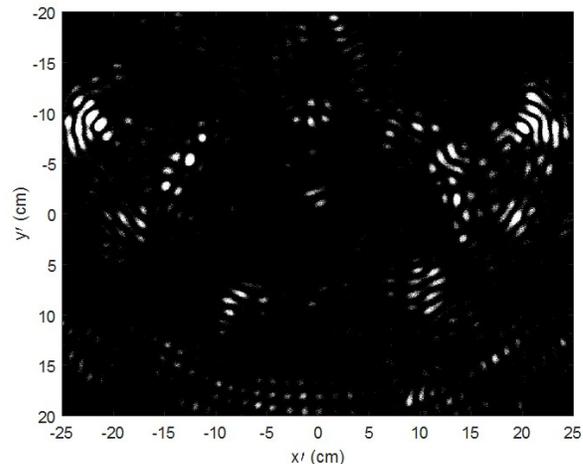

Fig. 8. Biological voxel model reconstruction with addition of SAR.

## VI. Conclusion

The method of applying SAR focusing on microwave reflection tomography data has been presented. An accuracy of within 5mm can be achieved for the process. The processing is stable and does not suffer from convergence problems like in inverse scattering. As it uses Fourier based image reconstruction, it is computationally efficient and fast.